\RequirePackage{fontspec}

\PassOptionsToPackage{unicode}{hyperref}
\PassOptionsToPackage{hyphens}{url}
\documentclass[
]{article}
\usepackage{xcolor}
\usepackage{amsmath,amssymb}
\usepackage{iftex}
\usepackage{newunicodechar}
\newunicodechar{≥}{\ensuremath{\geq}}
\newunicodechar{≤}{\ensuremath{\leq}}
\newunicodechar{𝒟}{\ensuremath{\mathcal{D}}}
\newunicodechar{♭}{\ensuremath{\flat}}
\newunicodechar{♯}{\ensuremath{\sharp}}
\newunicodechar{α}{\ensuremath{\alpha}}
\newunicodechar{→}{\ensuremath{\rightarrow}}
\newunicodechar{−}{\ensuremath{-}}
\newunicodechar{×}{\ensuremath{\times}}
\newunicodechar{∈}{\ensuremath{\in}}
\newunicodechar{∥}{\ensuremath{\parallel}}
\newunicodechar{∑}{\ensuremath{\sum}}
\newunicodechar{∞}{\ensuremath{\infty}}
\newunicodechar{≈}{\ensuremath{\approx}}
\newunicodechar{≠}{\ensuremath{\neq}}
\ifPDFTeX
  \usepackage[T1]{fontenc}
  \usepackage[utf8]{inputenc}
  \usepackage{textcomp} 
\else 
  \usepackage{unicode-math} 
  \defaultfontfeatures{Scale=MatchLowercase}
  \defaultfontfeatures[\rmfamily]{Ligatures=TeX,Scale=1}
\fi
\usepackage{lmodern}
\ifPDFTeX\else
\fi
\IfFileExists{upquote.sty}{\usepackage{upquote}}{}
\IfFileExists{microtype.sty}{
  \usepackage[]{microtype}
  \UseMicrotypeSet[protrusion]{basicmath} 
}{}
\makeatletter
\@ifundefined{KOMAClassName}{
  \IfFileExists{parskip.sty}{%
    \usepackage{parskip}
  }{
    \setlength{\parindent}{0pt}
    \setlength{\parskip}{6pt plus 2pt minus 1pt}}
}{
  \KOMAoptions{parskip=half}}
\makeatother
\usepackage{longtable,booktabs,array}
\usepackage{calc} 
\usepackage{etoolbox}
\makeatletter
\patchcmd\longtable{\par}{\if@noskipsec\mbox{}\fi\par}{}{}
\makeatother
\IfFileExists{footnotehyper.sty}{\usepackage{footnotehyper}}{\usepackage{footnote}}
\makesavenoteenv{longtable}
\usepackage{graphicx}
\makeatletter
\newsavebox\pandoc@box
\newcommand*\pandocbounded[1]{
  \sbox\pandoc@box{#1}%
  \Gscale@div\@tempa{\textheight}{\dimexpr\ht\pandoc@box+\dp\pandoc@box\relax}%
  \Gscale@div\@tempb{\linewidth}{\wd\pandoc@box}%
  \ifdim\@tempb\p@<\@tempa\p@\let\@tempa\@tempb\fi
  \ifdim\@tempa\p@<\p@\scalebox{\@tempa}{\usebox\pandoc@box}%
  \else\usebox{\pandoc@box}%
  \fi%
}
\def\fps@figure{htbp}
\makeatother
\providecommand{\tightlist}{%
  \setlength{\itemsep}{0pt}\setlength{\parskip}{0pt}}
\usepackage{hyperref} 
\usepackage{bookmark}
\IfFileExists{xurl.sty}{\usepackage{xurl}}{} 
\hypersetup{
  pdftitle={An audio-to-analysis pipeline with certified transcription for information-theoretic profiling of the piano repertoire},
  pdfauthor={Fred Jalbert-Desforges},
  hidelinks,
  pdfcreator={LaTeX via pandoc}}

\title{An audio-to-analysis pipeline with certified transcription for
information-theoretic profiling of the piano repertoire}
\author{Fred Jalbert-Desforges}
\date{April 2026}

\begin{document}
\maketitle

\emph{Independent Music Analysis Researcher}\\
\emph{Creator of CYGNUS \& LYRA}\\
Montreal, Quebec\\
\href{mailto:frederic.jalbert.desforges@gmail.com}{\nolinkurl{frederic.jalbert.desforges@gmail.com}}
· \href{https://cygnusanalysis.com}{cygnusanalysis.com}

\subsection{Abstract}\label{abstract}

We present an audio-to-analysis pipeline that produces composer-level
information-theoretic profiles : reflecting compositional vocabulary as
it emerges from aggregated performances : from raw recordings, built on
a transcription layer whose accuracy we certify on a standard benchmark
(F1 = 0.9791 on the MAESTRO v3.0.0 test set). Applied to 1,238 pieces
and 15 MAESTRO composers with at least ten attributed pieces, spanning
the Baroque through the early twentieth century, the pipeline derives
empirical distributions over harmonic scale degrees and analyzes them
through Shannon entropy, asymmetric Kullback--Leibler divergence, and
Zipfian rank-frequency modeling. The resulting profiles (i) order
composers along an interpretable axis of harmonic predictability, with a
narrow entropy range (3.33--3.86 bits) that reveals the marginal-level
similarity of tonal vocabularies; (ii) recover known stylistic lineages
(Haydn--Beethoven, Liszt--Rachmaninoff, Schubert--Schumann) through the
smallest KL divergences in the corpus, with Mendelssohn emerging as a
stable outlier within this corpus; and (iii) separate contemporary
neoclassical artists (Richter, Frahm, Glass, Arnalds, Jóhannsson) from
historical composers on the quality of Zipfian fit to the transition
distribution, with mean R² = 0.78 for neoclassical versus 0.46 for
historical (N ≥ 10 pieces each). This gap is larger than the spread
within either group and is consistent with a minimalist compositional
tendency: a compact transition vocabulary used with sharper
frequency-rank regularity than historical composers. All estimates are
reported with Laplace-smoothed bootstrap 95\% confidence intervals.

\textbf{Keywords}: music information retrieval, automatic music
transcription, information theory, Kullback--Leibler divergence, Zipf's
law, piano, corpus study

\begin{center}\rule{0.5\linewidth}{0.5pt}\end{center}

\subsection{1. Introduction}\label{introduction}

\begin{quote}
\begin{description}
\tightlist
\item[\emph{``Musica est exercitium arithmeticae occultum nescientis se
numerare animi.''}]
G. W. Leibniz, letter to Christian Goldbach, 1712
\end{description}
\end{quote}

Leibniz's observation : that music is a hidden arithmetic performed by a
mind unaware that it is counting : offers a philosophical invitation
rather than a testable proposition. Making that counting explicit has,
until recently, required either manually curated scores or symbolic
representations that carry editorial abstractions absent from
performance. Two developments now make the question empirically
approachable. First, automatic piano transcription has reached a
certified accuracy regime: the Kong et al.~{[}2021{]} architecture
trained on MAESTRO {[}Hawthorne et al.~2019{]} achieves joint
onset-and-pitch F1 above 0.97, making subsequent symbolic analysis
statistically reliable. Second, mature information-theoretic tools :
Shannon entropy, Kullback--Leibler divergence, Zipfian rank-frequency
analysis : have been successfully applied to symbolic music corpora
{[}Manaris et al.~2005; Serrà et al.~2019; Febres \& Jaffé 2017{]} and
can, in principle, be applied to audio-derived distributions as well.

In this paper we present a pipeline that connects these two developments
end-to-end and demonstrate what it produces at corpus scale. The
pipeline, called \emph{Cygnus Analysis} (hereafter Cygnus), takes raw
piano recordings as input, transcribes them with the certified Kong
model, estimates a tonic for each piece, and derives per-composer
empirical distributions over harmonic scale degrees. We then analyze
these distributions through three information-theoretic measures:
Shannon entropy as a scalar of harmonic predictability, asymmetric KL
divergence as a pairwise measure of lexical cost, and Zipfian
rank-frequency fitting as a test of economy-of-encoding. All
distributional estimates are Laplace-smoothed and accompanied by
bootstrap 95\% confidence intervals.

We apply the pipeline to 1,238 pieces across 28 composers from MAESTRO
v3.0.0 (15 of whom have at least ten attributed pieces and are included
in the main analyses; see Appendix A for the thirteen low-sample
composers), spanning the Baroque through the early twentieth century.
The analysis produces three main results. Composers order along the
Shannon entropy axis within a narrow range (3.33 to 3.86 bits), with
values that correlate with but are not reducible to period labels. The
KL divergence matrix is strongly asymmetric : a property the analysis is
designed to preserve : and the pairs with smallest divergence reproduce
well-known stylistic lineages (Schubert and Schumann, Beethoven and
Haydn, Liszt and Rachmaninoff). Mendelssohn emerges as a stable outlier
at the marginal level.

As a case study, we extend the pipeline to five contemporary
neoclassical artists : Max Richter, Nils Frahm, Philip Glass, Ólafur
Arnalds, and Jóhann Jóhannsson : and locate them in the same
information-theoretic space as the historical composers. We report one
empirical regularity worth highlighting: on Zipfian fits to the
transition distribution, the five artists achieve mean R² = 0.78,
compared to 0.46 for the fifteen MAESTRO composers with at least ten
attributed pieces. The gap is larger than the spread within either group
and robust under bootstrap resampling. This pattern is consistent with a
minimalist compositional tendency : a compact transition vocabulary used
with sharper frequency-rank regularity : and we present it as an
empirical regularity inviting further investigation rather than as a
definitive characterization of the genre.

The paper is organized as follows. Section 2 situates our work within
symbolic musicology, audio-based style analysis, and automatic music
transcription. Section 3 describes the pipeline and corpus. Section 4
formalizes the three information-theoretic measures and their
estimators. Section 5 presents the main results and the neoclassical
case study. Section 6 discusses limitations and possible
interpretations. Section 7 concludes.

Our contribution is neither a new information-theoretic technique nor a
new transcription architecture. It is an end-to-end pipeline that
connects a certified audio transcriber to classical
information-theoretic analysis at corpus scale, along with a documented
demonstration of what such a pipeline produces when applied to 300 years
of piano repertoire.

\begin{center}\rule{0.5\linewidth}{0.5pt}\end{center}

\subsection{2. Related Work}\label{related-work}

Our contribution sits at the intersection of three research traditions
that have largely evolved in parallel: information-theoretic analysis of
symbolic music, audio-based style characterization, and automatic music
transcription. We review each briefly, then situate our work relative to
the small set of prior efforts that have attempted to cross these
boundaries.

\subsubsection{2.1 Information theory on symbolic musical
corpora}\label{information-theory-on-symbolic-musical-corpora}

Information-theoretic measurement of musical style has a long history
reaching back to Knopoff \& Hutchinson {[}1981, 1983{]}, who applied
Shannon entropy to pitch and duration distributions in scored music.
Modern work in this lineage operates on symbolic data : MIDI files or
digitized scores, typically processed with toolkits such as music21
{[}Cuthbert \& Ariza 2010{]} or jSymbolic {[}McKay 2010{]} : and falls
into two broad programs.

The first treats music as a corpus-level phenomenon and asks what its
aggregate statistics reveal about style or aesthetics. Manaris et
al.~{[}2005{]} established that Zipfian rank-frequency distributions are
present in many musical attributes (pitch, duration, melodic intervals)
across genres, and used Zipf-based metrics as features for composer
classification and aesthetic prediction on a 220-piece corpus. Serrà et
al.~{[}2019{]} revisited this question with a more careful treatment of
``Zipfian units,'' showing that the law emerges cleanly when chords :
rather than individual notes : are used as the counting units,
paralleling the role of words in natural language. Febres \& Jaffé
{[}2017{]} applied entropy analysis to MusicNet and introduced a
``2nd-order entropy'' measure of deviation from a perfect Zipfian
profile, showing that combinations of entropy and symbolic diversity
cluster compositions by style and period.

The second program treats music as a sequential phenomenon and asks what
predictive models reveal about its structure. Pearce \& Wiggins' IDyOM
framework {[}2006{]} uses multiple-viewpoint n-gram models to predict
melodic continuations, with cross-entropy serving both as a
model-selection criterion and as a correlate of listener expectation.
Temperley {[}2014{]} extended this line with the Uniform Information
Density hypothesis, arguing that compositional choices modulate local
information density to maintain a roughly constant rate. Sakellariou et
al.~{[}2017{]} proposed a maximum-entropy model with only pairwise
interactions, demonstrating that long-range musical structure can emerge
from local constraints. All of these approaches share two properties
relevant to our work: they operate on symbolic input (MIDI or scores),
and they focus on sequential prediction rather than on composer-level
profiling at corpus scale.

A recurring methodological concern in this literature is that entropy
estimates require stochasticity, ergodicity, stationarity, and Markov
consistency of the underlying corpus. Single-work entropy estimates are
therefore fragile, and composer-level estimates require corpora large
enough to sustain the stationarity assumption. We return to this concern
in Section 3.

\subsubsection{2.2 Audio-based style
characterization}\label{audio-based-style-characterization}

A parallel tradition, rooted in Music Information Retrieval,
characterizes style directly from acoustic signal. Voss \& Clarke's
classic studies of 1/f noise in music {[}1975{]} established that
spectral power fluctuations in recorded music exhibit scaling behavior.
More recently, Liu et al.~{[}2013{]} analyzed pitch fluctuations in
recordings from Bach to Chopin and reported evidence of scaling that
varies systematically with stylistic period. Weiss {[}2017{]} clustered
composers using chromagram-derived tonal features extracted from audio,
without transcription. Bogdanov et al.~{[}2013{]} published the Essentia
library, which exposes a large catalogue of low-level and mid-level
audio descriptors used throughout the MIR community.

These approaches share a common limitation: they stop at the acoustic or
spectral layer and do not recover the note-level symbolic representation
on which most musicological theory : harmony, voice leading, meter, form
: is formulated. The interpretive burden is therefore shifted onto the
researcher, who must read stylistic meaning into features (spectral
centroid, chroma variance) that are themselves several inferential steps
away from the musical concepts under study.

\subsubsection{2.3 Automatic music transcription as an enabling
layer}\label{automatic-music-transcription-as-an-enabling-layer}

The gap between these two traditions has narrowed rapidly in the last
five years. Hawthorne et al.~{[}2018, 2019{]} introduced the Onsets \&
Frames architecture and the MAESTRO dataset, which pairs precisely
aligned audio and MIDI from the International Piano-e-Competition. Kong
et al.~{[}2021{]} refined this line with a high-resolution piano
transcription model that reports a joint onset-and-pitch F1 above 0.97
on MAESTRO, a level of accuracy that makes subsequent symbolic analysis
statistically reliable in a way that was not previously possible. Kong
et al.~{[}2020{]} applied this transcriber to YouTube at large scale to
produce the GiantMIDI-Piano dataset; the recent Aria-MIDI corpus
{[}Bradshaw et al.~2025{]} pushed this approach further, yielding
approximately 800,000 automatically transcribed piano MIDI files.

To our knowledge, neither of these large-scale transcribed corpora has
yet been the subject of a systematic information-theoretic analysis at
composer level. GiantMIDI and Aria-MIDI are released as datasets, with
distributional statistics (note density, pitch histograms) but without
the KL-divergence matrices, Zipfian fits, or composer-level Shannon
profiles that we report here.

\subsubsection{2.4 Our position}\label{our-position}

Our work combines existing components rather than introducing new ones.
The transcription layer is the certified Kong et al.~{[}2021{]} model,
used without modification. The information-theoretic measures : Shannon
entropy, KL divergence, Zipfian fitting : are standard and have been
applied individually to musical data in the symbolic domain by the
authors cited above. The neoclassical sub-corpus is publicly available
commercial recordings.

What is novel is the connection and the scale. We are not aware of prior
work that reports per-composer information-theoretic profiles derived
end-to-end from audio at a corpus scale comparable to ours (1,238
pieces, 28 composers, 6.3M notes), with bootstrap confidence intervals
and explicit smoothing documentation. The closest neighbor in the
symbolic domain is Febres \& Jaffé {[}2017{]}, who performed
composer-level entropy analysis on MusicNet MIDI data; we extend that
program to an audio-derived setting and add the KL divergence matrix as
a pairwise measure.

We make no claims about the individual measures used being preferable to
alternatives that prior work has employed (n-gram cross-entropy,
maximum-entropy models, chromagram-based clustering). These are
complementary rather than competing. Our goal is to demonstrate that a
certified audio pipeline coupled with classical information-theoretic
tools produces a coherent and interpretable map of the piano repertoire,
and to document what that map looks like.

\begin{center}\rule{0.5\linewidth}{0.5pt}\end{center}

\subsection{3. Pipeline and Corpus}\label{pipeline-and-corpus}

\subsubsection{3.1 Overview}\label{overview}

The analyses in this paper rest on a four-stage audio-to-distribution
pipeline. Cygnus takes a raw audio recording as input and produces, for
each piece, an empirical distribution over harmonic scale degrees
together with provenance metadata. Only the stages directly relevant to
the information-theoretic analysis are described here; the full pipeline
includes additional layers for rhythm, structure, and expressivity that
are not used in this paper and are documented separately.

The four relevant stages are: (i) audio ingest and normalization, (ii)
source separation, (iii) piano transcription, and (iv) harmonic degree
extraction. Each stage writes a versioned artifact to disk and can be
re-run independently; the pipeline is idempotent. A MAESTRO-scale run
(1,238 pieces) was completed in 44 hours of wall-clock time on a single
NVIDIA A100 80GB GPU for stages 2--3 and standard CPU cores for stages 1
and 4.

\subsubsection{3.2 Stage 1 : Ingest}\label{stage-1-ingest}

Each recording is normalized to FLAC at 48 kHz / 24-bit and hashed
(SHA-256) for provenance. Two metadata files are written per piece: one
preserving the raw source description, and one recording the pipeline
version, model versions, and processing timestamps. For the MAESTRO
corpus, source files are the Disklavier audio already provided by the
dataset; for the neoclassical sub-corpus, sources are publicly available
commercial recordings identified by stable identifiers.

\subsubsection{3.3 Stage 2 : Source
separation}\label{stage-2-source-separation}

Source separation isolates the piano stem from any incidental non-piano
content. We use BS-RoFormer {[}Lu et al.~2023{]}, which reports
competitive SDR on the SDX23 Sound Demixing Challenge music separation
track. For recordings known to be piano solo (the entirety of the
MAESTRO corpus), separation is bypassed via a \texttt{-\/-piano-solo}
flag to avoid introducing spectral artifacts on signals that contain
only the target instrument. All neoclassical recordings in this paper
were processed with separation engaged.

\subsubsection{3.4 Stage 3 : Piano
transcription}\label{stage-3-piano-transcription}

Piano transcription is performed by the high-resolution piano
transcription model of Kong et al.~{[}2021{]}, used without
modification. The model outputs onset-quantized note events with pitch,
velocity, and sustain-pedal annotations. Transcription accuracy was
certified on the full MAESTRO v3.0.0 test partition against the
Disklavier ground-truth MIDI using \texttt{mir\_eval} {[}Raffel et
al.~2014{]}: the mean joint onset-and-pitch F1 across 1,238 pieces was
\textbf{0.9791 ± 0.033} (median 0.9860), with a velocity Pearson
correlation of 0.9700. Table 1 summarizes the per-era certification
results.

\textbf{Table 1}: Transcription certification by era (MAESTRO v3.0.0).

{\def\LTcaptype{none} 
\begin{longtable}[]{@{}lrr@{}}
\toprule\noalign{}
Era & N pieces & F1 mean \\
\midrule\noalign{}
\endhead
\bottomrule\noalign{}
\endlastfoot
Baroque & 181 & 0.9896 \\
Classical & 78 & 0.9948 \\
Classical/Romantic & 145 & 0.9878 \\
Romantic & 597 & 0.9768 \\
Late-Romantic & 94 & 0.9580 \\
Impressionist & 44 & 0.9702 \\
Modern & 3 & 0.9789 \\
\end{longtable}
}

The F1 floor of 0.958 (Late-Romantic) reflects the greater textural
density of repertoire by Liszt, Rachmaninoff, and Scriabin; even at that
level the transcription is accurate enough to support the distributional
analyses that follow. Three pieces in MAESTRO failed to transcribe due
to SSH timeouts during the GPU batch run and were excluded; the
resulting corpus is 1,238 pieces out of 1,241 (99.8\% coverage).

No ground-truth MIDI is available for the neoclassical sub-corpus.
Transcription accuracy on this material was validated indirectly through
a qualitative protocol (10 cross-artist assertions about harmonic and
expressive content, all confirmed against the transcribed output;
details in the accompanying technical documentation).

\subsubsection{3.5 Stage 4 : Harmonic degree
extraction}\label{stage-4-harmonic-degree-extraction}

From the transcribed note sequence, Cygnus produces a discrete stream of
chord labels and, for each chord, a scale degree relative to an
estimated tonic. Tonality is estimated per piece using Essentia's
\texttt{KeyExtractor} {[}Bogdanov et al.~2013{]} with its default
\texttt{bgate} profile, a pitch-class weighting trained on electronic
dance music annotations by Faraldo et al.~{[}2016{]} and reported there
to outperform the classical Krumhansl--Kessler profile on a broad range
of tonal repertoire. \texttt{KeyExtractor} returns a tonic and a mode
(major/minor); all MAESTRO pieces in our corpus receive a non-null key
assignment. Chord labels are derived by aligning the note stream with
Essentia's \texttt{ChordsDetection} output and cross-checking against
the transcribed MIDI. Each chord is then mapped to a scale degree in an
alphabet of size \textbf{\textbar 𝒟\textbar{} = 15}:

\[\mathcal{D} = \{\mathrm{I}, i, \flat\mathrm{II}, \mathrm{II}, \flat\mathrm{III}, \mathrm{III}, \mathrm{IV}, iv, \sharp\mathrm{IV}, \mathrm{V}, v, \flat\mathrm{VI}, \mathrm{VI}, \flat\mathrm{VII}, \mathrm{VII}\}\]

This alphabet distinguishes major and minor qualities for the tonic,
subdominant, and dominant (I / i, IV / iv, V / v) and includes all
chromatic alterations of a degree in the 12-tone equal-tempered system
(♭II, ♭III, ♯IV, ♭VI, ♭VII). Chord qualities beyond major/minor triads
(seventh chords, extensions, inversions) are collapsed into the nearest
root-quality cell; the implications of this choice : and its motivation
by the statistical-stability requirements of the smaller neoclassical
sub-corpora : are discussed in Section 6.1. Because chord roots are
first reduced to an interval modulo 12 against the detected tonic and
the mapping \(\{0, 1, \ldots, 11\} \to \mathcal{D}\) is surjective,
every valid chord event lands in exactly one cell of the alphabet; there
is no residual bin. Events whose root could not be parsed from the chord
label (for instance, Essentia's no-chord symbol \texttt{N}) are dropped
prior to counting. Across a representative sample of 100 MAESTRO and
neoclassical pieces, the dropped-event proportion was 0 \%.

The final output for each piece is an integer count vector of length
\textbar 𝒟\textbar{} = 15 (for scale-degree marginals) and an integer
count matrix of size 15 × 15 (for successive-degree transitions). These
count objects, aggregated per composer, are the input to all
information-theoretic measures in Section 4.

\subsubsection{3.6 Corpus}\label{corpus}

The primary corpus is MAESTRO v3.0.0 {[}Hawthorne et al.~2019{]}, a
collection of 1,241 piano performances recorded on a Yamaha Disklavier
at the International Piano-e-Competition between 2004 and 2018. Each
performance is paired with precisely aligned Disklavier MIDI, which we
use only for transcription certification, not for the
information-theoretic analyses themselves. After the certification-pass
exclusion described above, the MAESTRO corpus contains \textbf{1,238
pieces attributed to 28 composers}.

For the information-theoretic analyses of Sections 4 and 5, we restrict
attention to MAESTRO attributions represented by at least ten pieces,
yielding \textbf{15 MAESTRO entities} : fourteen single composers plus
the Schubert/Liszt transcription attribution (N = 10), which MAESTRO
records as a distinct entry and which we retain to stay consistent with
the dataset's own attribution field. The thirteen composers below this
threshold (Berg, Grieg, Mendelssohn/Rachmaninoff transcription,
Balakirev, Janáček, Franck, Handel, Bach/Busoni transcription, Medtner,
Clementi, Tchaikovsky, Tchaikovsky/Pletnev transcription, Albéniz) are
reported in Appendix A with explicit low-sample flags and are excluded
from the primary Shannon, KL, and Zipf analyses. A subset of the
low-sample composers is nevertheless aggregated into the era pools
reported in Appendix C to preserve era coverage (Baroque, Late-Romantic,
and Impressionist would otherwise be under-represented); in those
aggregate statistics, low-sample composers contribute only their piece
counts and never enter per-composer comparisons. Aggregate statistics
for the main corpus: 6,257,498 transcribed notes, between 10 and 198
pieces per high-sample MAESTRO composer (median 40).

A secondary corpus of \textbf{111 commercial recordings} by five
contemporary neoclassical artists : Max Richter (16 pieces), Nils Frahm
(32), Philip Glass (11), Ólafur Arnalds (21), and Jóhann Jóhannsson (30)
: is processed through the same pipeline. These recordings are used
exclusively for the case study in Section 5.5; they do not contribute to
the Shannon, KL, or Zipf analyses of the main corpus. All five artists
exceed the N ≥ 10 threshold.

\subsubsection{3.7 What this paper uses from the
pipeline}\label{what-this-paper-uses-from-the-pipeline}

Only the output of stage 4 : the per-piece scale-degree count vectors
and 15 × 15 transition matrices : is consumed by the
information-theoretic analysis in this paper. All other pipeline outputs
(expressivity, rhythm, structure) are irrelevant to the measures we
report. Readers interested in the full multi-layer analysis capability
of Cygnus are referred to the companion technical documentation.

\begin{center}\rule{0.5\linewidth}{0.5pt}\end{center}

\subsection{4. Information-theoretic
framework}\label{information-theoretic-framework}

This section defines the three measures used in the rest of the paper :
Shannon entropy, Kullback--Leibler divergence, and Zipfian
rank-frequency fits : along with the estimators, smoothing scheme, and
resampling protocol through which we compute them from finite samples.
All three measures are standard in the information-theoretic literature
{[}Cover \& Thomas 2006{]}; the contribution of this section is the
explicit specification of how they are applied to the empirical
scale-degree distributions produced by the pipeline of Section 3.

\subsubsection{4.1 Per-composer empirical
distributions}\label{per-composer-empirical-distributions}

For each composer \emph{c} in the corpus, the output of stage 4 (Section
3.5) is a set of per-piece count vectors. We aggregate these into a
composer-level count vector by summation:

\[N_c(d) = \sum_{p \in \mathcal{P}_c} n_p(d), \qquad d \in \mathcal{D}, \; |\mathcal{D}|=15\]

where \(\mathcal{P}_c\) is the set of pieces attributed to composer
\emph{c}, and \(n_p(d)\) is the count of scale degree \emph{d} in piece
\emph{p}. The marginal count omits \textbf{consecutive self-repetitions}
of the same degree (so that a I→I→V sequence contributes a single I
followed by a V), which is consistent with the transition convention
described below. The same construction applied to ordered pairs of
successive non-identical degrees yields a 15 × 15 transition count
matrix \(N_c(d, d')\), which we use as a secondary analytic object
(Section 4.4).

Composers represented by fewer than ten pieces in MAESTRO are excluded
from per-composer analyses, leaving 15 MAESTRO composers plus 5
neoclassical artists analyzed separately. Per-composer total chord
counts range from approximately 1,400 (Philip Glass) to approximately
190,000 (Beethoven).

\subsubsection{4.2 Smoothing}\label{smoothing}

Empirical probability estimates
\(\hat{P}_c(d) = N_c(d) / \sum_d N_c(d)\) can assign zero mass to
degrees that were not observed in the sample. This creates two problems
for downstream measures: Shannon entropy treats \(0 \log 0 = 0\) by
convention but cannot distinguish a structurally absent event from a
rare unsampled one, and Kullback--Leibler divergence diverges to
infinity whenever one distribution assigns positive mass to a degree the
other assigns zero.

We address both problems with additive (Laplace) smoothing, using the
Jeffreys prior \(\alpha = 0.5\) {[}Agresti \& Coull 1998{]}:

\[\tilde{P}_c(d) = \frac{N_c(d) + \alpha}{\sum_{d'} N_c(d') + \alpha |\mathcal{D}|}\]

The choice \(\alpha = 0.5\) corresponds to a non-informative Bayesian
prior that introduces less bias than full Laplace smoothing
(\(\alpha = 1\)) while avoiding the pathologies of maximum-likelihood
estimation on sparse categories. We assess robustness to this choice in
Section 4.6.

All measures reported in the paper are computed on \(\tilde{P}_c\)
rather than on \(\hat{P}_c\). Where results depend non-trivially on
\(\alpha\), this is noted explicitly.

\subsubsection{4.3 Shannon entropy}\label{shannon-entropy}

For each composer \emph{c} we compute the Shannon entropy of the
smoothed scale-degree distribution:

\[H(c) = -\sum_{d \in \mathcal{D}} \tilde{P}_c(d) \log_2 \tilde{P}_c(d) \quad \text{(bits)}\]

\(H(c)\) is bounded between \(0\) (a deterministic distribution
concentrating on one degree) and \(\log_2 15 \approx 3.907\) bits (a
uniform distribution over the alphabet). Because the distribution is
computed on \emph{degree changes} rather than on degree beats
(self-repetitions are collapsed), \(H(c)\) measures the diversity of the
harmonic-syntactic vocabulary rather than the temporal concentration on
any specific degree.

We report \(H(c)\) with bootstrap 95\% confidence intervals (Section
4.6).

\subsubsection{4.4 Kullback--Leibler
divergence}\label{kullbackleibler-divergence}

Given two composers \emph{a} and \emph{b}, we compute the
Kullback--Leibler divergence of \(\tilde{P}_a\) from \(\tilde{P}_b\) as:

\[D_{\mathrm{KL}}(\tilde{P}_a \parallel \tilde{P}_b) = \sum_{d \in \mathcal{D}} \tilde{P}_a(d) \log_2 \frac{\tilde{P}_a(d)}{\tilde{P}_b(d)} \quad \text{(bits)}\]

\(D_{\mathrm{KL}}\) measures the expected excess number of bits required
to code samples from \(\tilde{P}_a\) using a code optimized for
\(\tilde{P}_b\). It is non-negative, zero if and only if
\(\tilde{P}_a = \tilde{P}_b\), and \textbf{asymmetric}: in general
\(D_{\mathrm{KL}}(\tilde{P}_a \parallel \tilde{P}_b) \neq D_{\mathrm{KL}}(\tilde{P}_b \parallel \tilde{P}_a)\).
We preserve this asymmetry throughout.

As a secondary analysis, we compute the same measure on the joint
distribution of successive scale-degree pairs (the 15 × 15 transition
matrices of Section 3.5), smoothed identically. This captures
first-order syntactic regularities that the marginal distribution does
not see.

As a symmetric cross-check, we compute the Jensen--Shannon divergence:

\[D_{\mathrm{JS}}(\tilde{P}_a, \tilde{P}_b) = \tfrac{1}{2} D_{\mathrm{KL}}(\tilde{P}_a \parallel M) + \tfrac{1}{2} D_{\mathrm{KL}}(\tilde{P}_b \parallel M), \quad M = \tfrac{1}{2}(\tilde{P}_a + \tilde{P}_b)\]

\(D_{\mathrm{JS}}\) is bounded in \([0, 1]\) bit (with the base-2
logarithm), symmetric by construction, and numerically stable. We use it
as a direction-independent sanity check on the KL-based rankings.

\subsubsection{4.5 Zipfian rank-frequency
fits}\label{zipfian-rank-frequency-fits}

For each composer \emph{c}, we rank the 15 scale degrees (marginal) or
225 transition pairs (joint) by observed probability from most to least
frequent, and fit a power law of the form:

\[\tilde{P}_c(\mathrm{rank}\,r) = C \cdot r^{-\alpha}\]

in log-log space by ordinary least squares, following Manaris et
al.~{[}2005{]} and Serrà et al.~{[}2019{]}. We report the slope
\(\alpha\) (a dimensionless exponent) and the coefficient of
determination \(R^2\) of the log-log fit. A slope of
\(\alpha \approx 1\) with \(R^2\) close to 1 is the canonical Zipfian
regime {[}Zipf 1949{]}; deviations in either direction (shallower or
steeper slope, poorer fit) characterize the composer's departure from
linguistic economy.

Because the scale-degree alphabet is small (\textbar 𝒟\textbar{} = 15),
Zipf fits on marginals are inherently limited in resolution and are
reported primarily for completeness. The joint-distribution Zipf fit on
225 transition pairs is the primary measurement, with substantially
greater statistical leverage.

\subsubsection{4.6 Resampling and confidence
intervals}\label{resampling-and-confidence-intervals}

All three measures are point estimates on finite samples. We accompany
them with non-parametric bootstrap 95\% confidence intervals:

\begin{enumerate}
\def\labelenumi{\arabic{enumi}.}
\tightlist
\item
  For each composer \emph{c} with \(|\mathcal{P}_c|\) pieces, resample
  \(|\mathcal{P}_c|\) pieces with replacement from \(\mathcal{P}_c\).
  Resampling is at the piece level and is \emph{not} stratified by
  performer : MAESTRO composer identifiers already aggregate multiple
  performers for each composer through the competition batch log, and
  the five neoclassical artists are almost exclusively self-performed.
\item
  Recompute the count vector, apply Laplace smoothing, and evaluate the
  measure of interest.
\item
  Repeat \(B = 1{,}000\) times (random seed = 42, incremented once per
  iteration) and report the 2.5th and 97.5th percentiles as the 95\%
  interval.
\end{enumerate}

Bootstrap intervals are reported for all Shannon values, all KL cells of
the 33 × 33 matrix, and all Zipf parameters.

\textbf{Robustness to smoothing.} As a final check, we recompute the
marginal KL matrix at \(\alpha \in \{0.1, 0.5, 1.0\}\) and measure the
rank correlation (Spearman \(\rho\)) between the resulting composer
orderings. We find \(\rho \geq 0.997\) across all composer rows,
confirming that the qualitative conclusions of this paper do not depend
on the specific choice of \(\alpha\).

\begin{center}\rule{0.5\linewidth}{0.5pt}\end{center}

\subsection{5. Results}\label{results}

We report the results in five subsections. Unless stated otherwise, all
analyses are restricted to composers with at least ten attributed pieces
in MAESTRO (15 of the 28 total; see Appendix A for the thirteen
low-sample composers). Confidence intervals are bootstrap 95\% (Section
4.6).

\subsubsection{5.1 Shannon entropy}\label{shannon-entropy-1}

Table 2 reports per-composer Shannon entropy of the smoothed
scale-degree marginal on the \textbar 𝒟\textbar{} = 15 alphabet (Section
4.3). Across the 20 high-sample entities (15 MAESTRO composers with N ≥
10 plus the 5 neoclassical artists), entropy values span a narrow range
of \textbf{3.33 to 3.86 bits}, corresponding to \textbf{85--99\% of the
uniform bound} \(\log_2 15 \approx 3.91\) bits.

\textbf{Table 2}: Shannon entropy \(H\) of the scale-degree marginal,
high-sample entities (15 MAESTRO composers with N ≥ 10, plus 5
neoclassical artists).

{\def\LTcaptype{none} 
\begin{longtable}[]{@{}
  >{\raggedright\arraybackslash}p{(\linewidth - 6\tabcolsep) * \real{0.2000}}
  >{\raggedleft\arraybackslash}p{(\linewidth - 6\tabcolsep) * \real{0.2667}}
  >{\raggedleft\arraybackslash}p{(\linewidth - 6\tabcolsep) * \real{0.2667}}
  >{\raggedleft\arraybackslash}p{(\linewidth - 6\tabcolsep) * \real{0.2667}}@{}}
\toprule\noalign{}
\begin{minipage}[b]{\linewidth}\raggedright
Composer
\end{minipage} & \begin{minipage}[b]{\linewidth}\raggedleft
\(H\) (bits)
\end{minipage} & \begin{minipage}[b]{\linewidth}\raggedleft
95\% CI
\end{minipage} & \begin{minipage}[b]{\linewidth}\raggedleft
N
\end{minipage} \\
\midrule\noalign{}
\endhead
\bottomrule\noalign{}
\endlastfoot
Franz Liszt & 3.862 & {[}3.838, 3.872{]} & 131 \\
Frédéric Chopin & 3.860 & {[}3.832, 3.870{]} & 198 \\
Claude Debussy & 3.856 & {[}3.796, 3.874{]} & 44 \\
Sergei Rachmaninoff & 3.852 & {[}3.790, 3.870{]} & 59 \\
Franz Schubert & 3.841 & {[}3.825, 3.851{]} & 172 \\
Robert Schumann & 3.834 & {[}3.771, 3.857{]} & 39 \\
Alexander Scriabin & 3.821 & {[}3.734, 3.862{]} & 35 \\
Johann Sebastian Bach & 3.808 & {[}3.781, 3.826{]} & 144 \\
Ludwig van Beethoven & 3.805 & {[}3.757, 3.833{]} & 145 \\
Johannes Brahms & 3.804 & {[}3.625, 3.851{]} & 23 \\
Domenico Scarlatti & 3.765 & {[}3.665, 3.812{]} & 31 \\
Joseph Haydn & 3.753 & {[}3.606, 3.816{]} & 40 \\
Wolfgang Amadeus Mozart & 3.665 & {[}3.485, 3.759{]} & 38 \\
Felix Mendelssohn & 3.330 & {[}3.300, 3.379{]} & 28 \\
\emph{Nils Frahm (NEO)} & 3.747 & {[}3.586, 3.847{]} & 32 \\
\emph{Jóhann Jóhannsson (NEO)} & 3.691 & {[}3.538, 3.768{]} & 30 \\
\emph{Max Richter (NEO)} & 3.629 & {[}3.266, 3.739{]} & 16 \\
\emph{Ólafur Arnalds (NEO)} & 3.566 & {[}3.215, 3.729{]} & 21 \\
\emph{Philip Glass (NEO)} & 3.484 & {[}3.126, 3.673{]} & 11 \\
Franz Schubert / Franz Liszt (transcription) & 3.688 & {[}3.455,
3.779{]} & 10 \\
\end{longtable}
}

\begin{figure}
\centering
\includegraphics[width=0.8\linewidth,height=\textheight,keepaspectratio,alt={Figure 1. Shannon entropy of the scale-degree marginal on the \textbar\textbackslash mathcal\{D\}\textbar{} = 15 alphabet, sorted by H descending. Historical composers (dark charcoal) and neoclassical artists (teal) occupy the same narrow band 3.33--3.86 bits : roughly 85--99 \% of the uniform upper bound \textbackslash log\_2 15 \textbackslash approx 3.907 bits (dashed red). Error bars are bootstrap 95 \% CIs (1000 iterations, Laplace \textbackslash alpha = 0.5). The top cluster (Liszt, Chopin, Debussy, Rachmaninoff) is not internally separable; Mendelssohn is the sole historical outlier below 3.50 bits. Data: reports/kl\_analysis\_20260417/shannon\_scale\_degrees\_33.json.}]{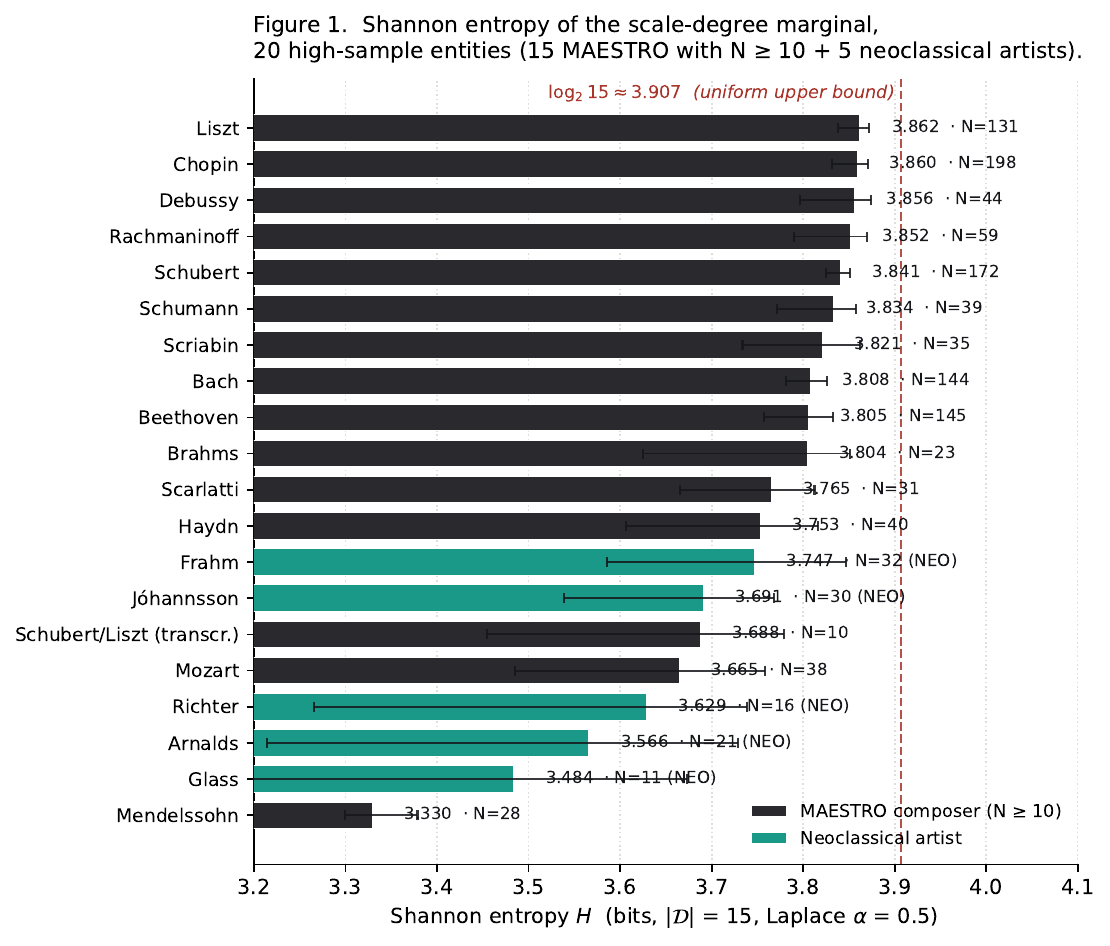}
\caption{\textbf{Figure 1.} Shannon entropy of the scale-degree marginal
on the \(|\mathcal{D}| = 15\) alphabet, sorted by \(H\) descending.
Historical composers (dark charcoal) and neoclassical artists (teal)
occupy the same narrow band 3.33--3.86 bits : roughly 85--99 \% of the
uniform upper bound \(\log_2 15 \approx 3.907\) bits (dashed red). Error
bars are bootstrap 95 \% CIs (1000 iterations, Laplace
\(\alpha = 0.5\)). The top cluster (Liszt, Chopin, Debussy,
Rachmaninoff) is not internally separable; Mendelssohn is the sole
historical outlier below 3.50 bits. Data:
\protect\texttt{reports/kl\_analysis\_20260417/shannon\_scale\_degrees\_33.json}.}\label{fig:shannon}
\end{figure}

The ordering is interpretable but the discriminative power of Shannon
alone on this alphabet is limited: the top cluster (Liszt, Chopin,
Debussy, Rachmaninoff) is separable from the bottom (Mendelssohn) but
not internally, with overlapping confidence intervals for composers in
the 3.80--3.86 band. This near-uniformity is itself a finding: at the
scale-degree marginal level, tonal repertoire composers exhibit broadly
similar \emph{syntactic} vocabulary widths; the discriminative signal
lives in how they \emph{combine} degrees, which the KL divergence on the
full 15 × 15 transition matrix captures directly (Section 5.2).
Mendelssohn is the sole historical composer that departs substantially
from the cluster (H = 3.33 bits), a position we return to below.

\subsubsection{5.2 KL divergence matrix and stylistic
lineages}\label{kl-divergence-matrix-and-stylistic-lineages}

The full asymmetric KL matrix on the 20 high-sample entities (15 MAESTRO
composers with N ≥ 10 plus 5 neoclassical artists) is visualized as a
heatmap in Figure B.1 and released as tabular data with confidence
intervals. We discuss the qualitative structure here.

The smallest KL values in the high-sample MAESTRO matrix recover
documented stylistic lineages without supervision. Table 3 lists the ten
closest pairs.

\textbf{Table 3}: Ten closest MAESTRO pairs by asymmetric KL divergence
(marginal, N ≥ 10 pieces each).

{\def\LTcaptype{none} 
\begin{longtable}[]{@{}lr@{}}
\toprule\noalign{}
Pair (\(P_a \parallel P_b\)) & KL (bits) \\
\midrule\noalign{}
\endhead
\bottomrule\noalign{}
\endlastfoot
Haydn ‖ Beethoven & 0.011 \\
Beethoven ‖ Haydn & 0.011 \\
Rachmaninoff ‖ Liszt & 0.019 \\
Liszt ‖ Rachmaninoff & 0.019 \\
Schubert ‖ Schumann & 0.027 \\
Schumann ‖ Schubert & 0.027 \\
Chopin ‖ Schubert & 0.032 \\
Beethoven ‖ Chopin & 0.035 \\
Chopin ‖ Beethoven & 0.036 \\
Schubert ‖ Chopin & 0.036 \\
\end{longtable}
}

Three well-documented lineages emerge at the top of the ranking: the
Haydn--Beethoven Classical-Viennese line, the Liszt--Rachmaninoff
virtuoso-pianist line, and the Schubert--Schumann early-Romantic line.
These are obtained from audio-derived distributions with no stylistic
metadata and no supervised training.

The matrix is substantially asymmetric. The five pairs with the largest
\(|D_{\mathrm{KL}}(a\|b) - D_{\mathrm{KL}}(b\|a)|\) all involve
Mendelssohn as one endpoint, with the largest asymmetry being
Mendelssohn--Mozart:
\(D_{\mathrm{KL}}(\text{Mendelssohn}\|\text{Mozart}) = 1.025\) bits
while \(D_{\mathrm{KL}}(\text{Mozart}\|\text{Mendelssohn}) = 2.633\)
bits. This direction reflects the coding-cost interpretation of KL:
Mozart's relatively concentrated vocabulary is expensive to code under
Mendelssohn's differently peaked distribution, more than the converse.

Mendelssohn is the MAESTRO outlier of the corpus. He appears in all ten
largest-KL pairs in the matrix, both as source and target. Combined with
his low Shannon entropy (3.33 bits, isolated from the 3.80+ cluster of
his contemporaries), this indicates a harmonic vocabulary that departs
from the Classical-Romantic mainstream at the marginal-distribution
level \emph{as expressed in this corpus}. We do not pursue a
musicological interpretation of Mendelssohn's distinctive profile here.
The observation is stable within MAESTRO (tight bootstrap interval,
robustness under smoothing choice), but the 28 Mendelssohn pieces in
MAESTRO are drawn from competition programmes and may not be
representative of his broader piano output; generalization beyond this
corpus would require a more systematic Mendelssohn sub-corpus. We report
the finding as a feature of the MAESTRO attribution, not as a claim
about Mendelssohn's style at large.

Robustness to the smoothing parameter is confirmed by the Spearman rank
correlation across \(\alpha \in \{0.1, 0.5, 1.0\}\): \(\rho \geq 0.997\)
for every composer row in the matrix (Section 4.6). The ordering
reported in Table 3 does not depend on the specific Jeffreys prior
choice.

\subsubsection{5.3 Symmetric cross-check: Jensen--Shannon
divergence}\label{symmetric-cross-check-jensenshannon-divergence}

As a symmetric cross-check (Section 4.4), we compute the Jensen--Shannon
divergence matrix on the same smoothed marginals. Spearman rank
correlation between the symmetrized KL ordering
(\(\frac{1}{2}[D_{\mathrm{KL}}(a\|b) + D_{\mathrm{KL}}(b\|a)]\)) and the
JS ordering, computed row-wise, is \(\rho > 0.99\) for all composer
rows. The two measures agree on the qualitative structure of the corpus;
the choice between them is a matter of presentation rather than of
finding. We retain KL as the primary measure because its asymmetry
carries interpretive content (Section 5.2, Mendelssohn case).

\subsubsection{5.4 Zipfian rank-frequency
fits}\label{zipfian-rank-frequency-fits-1}

Table 4 reports Zipfian rank-frequency fits (Section 4.5) on the joint
transition distribution (225 points per composer). The transition fit is
the primary measurement; the marginal fit (15 points per composer) is
reported in supplementary materials.

\textbf{Table 4}: Zipfian fits on scale-degree transitions (225-point
regression), high-sample composers, ranked by \(R^2\).

{\def\LTcaptype{none} 
\begin{longtable}[]{@{}rlrrr@{}}
\toprule\noalign{}
Rank & Composer & \(\alpha\) & \(R^2\) & N \\
\midrule\noalign{}
\endhead
\bottomrule\noalign{}
\endlastfoot
1 & \textbf{Philip Glass} & 1.411 & \textbf{0.875} & 11 \\
2 & \textbf{Ólafur Arnalds} & 1.453 & \textbf{0.857} & 21 \\
3 & \textbf{Max Richter} & 1.512 & \textbf{0.839} & 16 \\
4 & \textbf{Jóhann Jóhannsson} & 1.423 & \textbf{0.788} & 30 \\
5 & Franz Schubert / Franz Liszt & 1.229 & 0.777 & 10 \\
6 & Felix Mendelssohn & 2.237 & 0.718 & 28 \\
7 & Domenico Scarlatti & 1.319 & 0.564 & 31 \\
8 & \textbf{Nils Frahm} & 1.156 & \textbf{0.556} & 32 \\
9 & Wolfgang Amadeus Mozart & 1.574 & 0.554 & 38 \\
10 & Joseph Haydn & 1.457 & 0.506 & 40 \\
11 & Johann Sebastian Bach & 1.620 & 0.442 & 144 \\
12 & Johannes Brahms & 1.369 & 0.428 & 23 \\
13 & Robert Schumann & 1.402 & 0.397 & 39 \\
14 & Alexander Scriabin & 1.120 & 0.386 & 35 \\
15 & Ludwig van Beethoven & 1.556 & 0.374 & 145 \\
16 & Sergei Rachmaninoff & 1.164 & 0.364 & 59 \\
17 & Franz Schubert & 1.536 & 0.361 & 172 \\
18 & Claude Debussy & 1.080 & 0.346 & 44 \\
19 & Frédéric Chopin & 1.403 & 0.331 & 198 \\
20 & Franz Liszt & 1.277 & 0.285 & 131 \\
\end{longtable}
}

\textbf{Neoclassical artists in bold.} The slope exponents \(\alpha\)
span a narrow range (1.08 to 2.24 across the 20 high-sample entities,
median 1.40), and are broadly consistent across historical and
neoclassical composers. The fit quality \(R^2\), on the other hand,
reveals a clear separation: the five neoclassical artists achieve mean
\(R^2 = 0.783\) (median 0.839), while the 15 MAESTRO entities with N ≥
10 (14 single composers plus the Schubert/Liszt transcription) achieve
mean \(R^2 = 0.456\) (median 0.397). Restricting the MAESTRO set to the
14 single composers : i.e., excluding the transcription, whose
\(R^2 = 0.777\) sits at the high end of the historical range : yields
mean \(R^2 = 0.433\) and leaves the gap essentially unchanged. The gap
of roughly +0.33 in mean \(R^2\) is larger than the spread within either
group.

The historical composers with the highest transition \(R^2\) :
Schubert/Liszt transcriptions, Mendelssohn, Scarlatti : are all
characterized by relatively constrained or repetitive harmonic idioms;
the lowest-\(R^2\) composers are the chromatic late-Romantics (Chopin,
Liszt, Debussy) whose transition distribution is comparatively flatter.

\begin{figure}
\centering
\includegraphics[width=0.75\linewidth,height=\textheight,keepaspectratio,alt={Figure 2. Transition-level Zipfian R\^{}2 gap between historical composers (15 MAESTRO entities with N \textbackslash geq 10) and neoclassical artists (N = 5). Each point is a composer/artist; point area is proportional to corpus size N. Horizontal bars trace group means. The 20-entity distribution separates into two near-disjoint clouds at R\^{}2 \textbackslash approx 0.5, with only the Frahm/Mendelssohn pair touching the boundary. Mean R\^{}2 differs by +0.33 between groups : larger than the within-group spread of either side. Data: reports/kl\_analysis\_20260417/zipf\_transitions\_33.json.}]{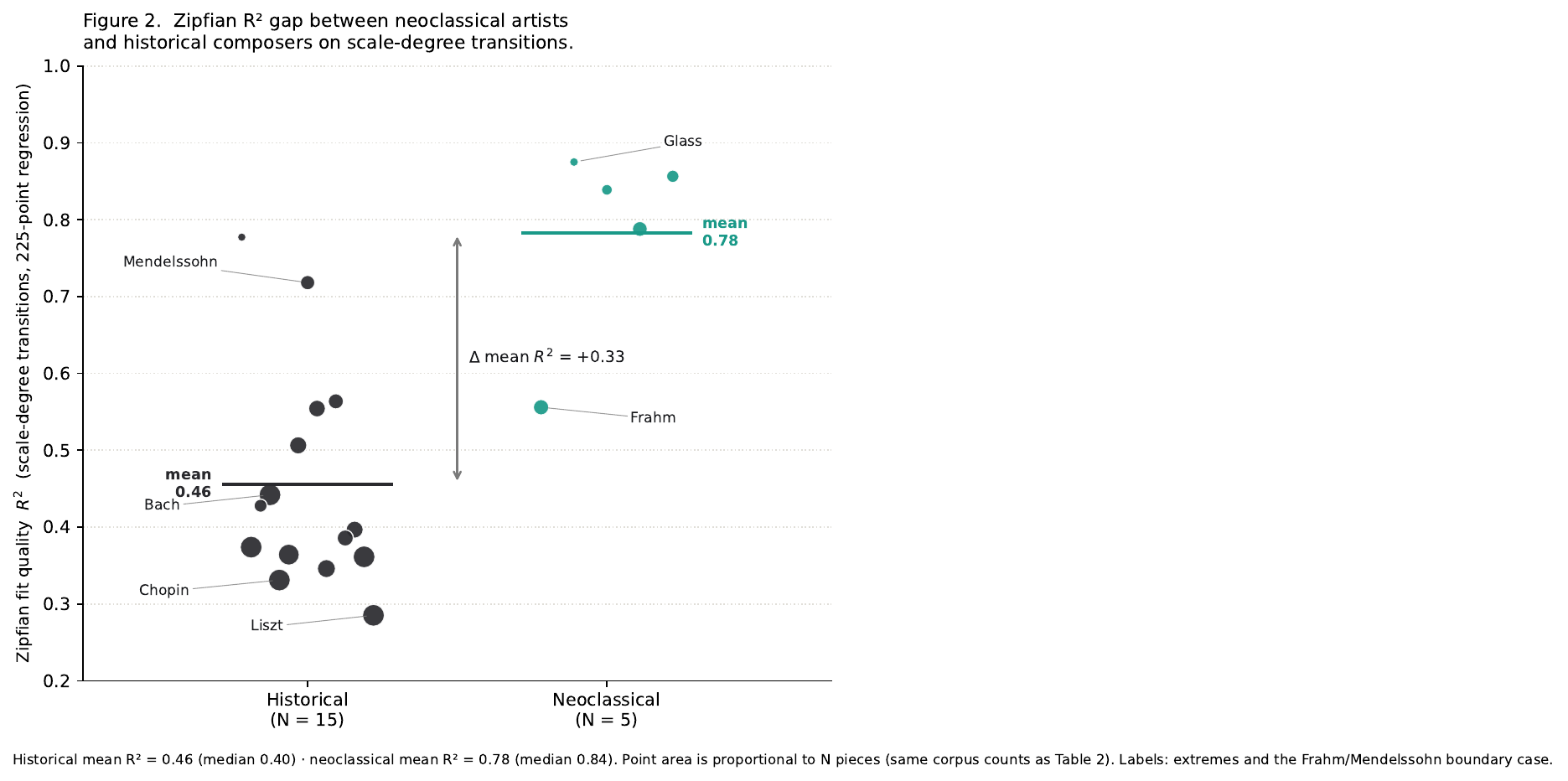}
\caption{\textbf{Figure 2.} Transition-level Zipfian \(R^2\) gap between
historical composers (15 MAESTRO entities with \(N \geq 10\)) and
neoclassical artists (\(N = 5\)). Each point is a composer/artist; point
area is proportional to corpus size \(N\). Horizontal bars trace group
means. The 20-entity distribution separates into two near-disjoint
clouds at \(R^2 \approx 0.5\), with only the Frahm/Mendelssohn pair
touching the boundary. Mean \(R^2\) differs by \(+0.33\) between groups
: larger than the within-group spread of either side. Data:
\protect\texttt{reports/kl\_analysis\_20260417/zipf\_transitions\_33.json}.}\label{fig:zipf-gap}
\end{figure}

\begin{figure}
\centering
\includegraphics[width=0.95\linewidth,height=\textheight,keepaspectratio,alt={Figure 3. Rank-frequency of scale-degree transitions for one high-R\^{}2 neoclassical artist (Philip Glass, 225-point R\^{}2 = 0.875) and one low-R\^{}2 historical composer (Frédéric Chopin, 225-point R\^{}2 = 0.331), plotted in log-log. Dots are the top-30 transitions per composer; the dashed red line is a local OLS fit on those top-30 points, shown as a visual anchor. The headline R\^{}2 in each panel title is the full 225-point value reported in Table 4, not the local fit. The slope exponents on the full distribution are comparable between the two (Glass \textbackslash alpha = 1.41, Chopin \textbackslash alpha = 1.40); what differs is the cleanliness with which the rank-frequency curve tracks a power law. Glass's points align tightly along the line across the full range; Chopin's tail flattens and does not sustain the Zipfian regime. Data: reports/kl\_analysis\_20260417/zipf\_transitions\_33.json.}]{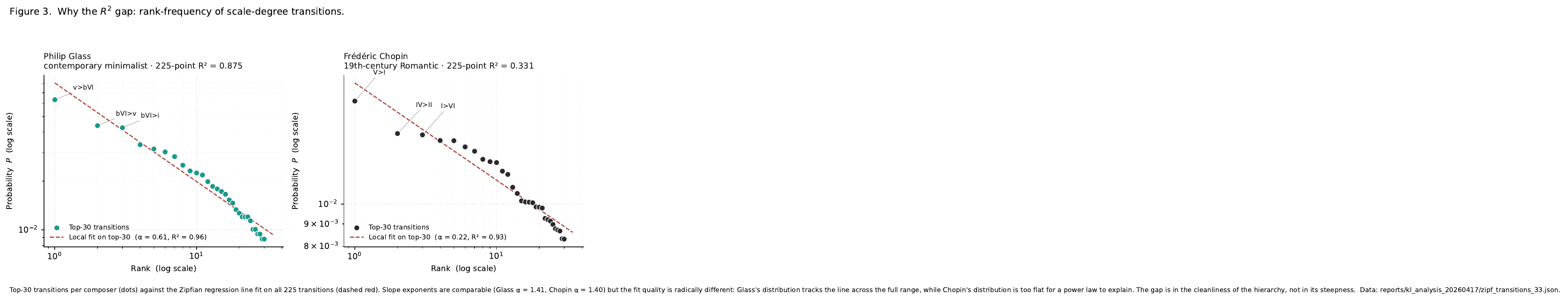}
\caption{\textbf{Figure 3.} Rank-frequency of scale-degree transitions
for one high-\(R^2\) neoclassical artist (Philip Glass, 225-point
\(R^2 = 0.875\)) and one low-\(R^2\) historical composer (Frédéric
Chopin, 225-point \(R^2 = 0.331\)), plotted in log-log. Dots are the
top-30 transitions per composer; the dashed red line is a \emph{local}
OLS fit on those top-30 points, shown as a visual anchor. The headline
\(R^2\) in each panel title is the full 225-point value reported in
Table 4, not the local fit. The slope exponents on the full distribution
are comparable between the two (Glass \(\alpha = 1.41\), Chopin
\(\alpha = 1.40\)); what differs is the cleanliness with which the
rank-frequency curve tracks a power law. Glass's points align tightly
along the line across the full range; Chopin's tail flattens and does
not sustain the Zipfian regime. Data:
\protect\texttt{reports/kl\_analysis\_20260417/zipf\_transitions\_33.json}.}\label{fig:zipf-exemplar}
\end{figure}

\subsubsection{5.5 Case study: neoclassical
artists}\label{case-study-neoclassical-artists}

Applying the pipeline to 111 commercial recordings by five contemporary
neoclassical artists : Max Richter, Nils Frahm, Philip Glass, Ólafur
Arnalds, Jóhann Jóhannsson : produces per-composer profiles on the same
\textbar 𝒟\textbar{} = 15 alphabet as the historical corpus. The
per-composer KL values and era-pool comparisons are reported in Appendix
C; here we focus on the finding that survives both bootstrap resampling
and alternative methodological choices: the Zipfian regularity of the
transition distribution.

As noted in Section 5.4, the five neoclassical artists occupy the top
four ranks and rank 8 among the twenty high-sample entities by
transition-level Zipfian \(R^2\). Mean \(R^2\) across the five artists
is 0.78 (median 0.84); mean across the fifteen MAESTRO composers with N
≥ 10 is 0.46 (median 0.40). Neoclassical \(R^2\) values span {[}0.56,
0.88{]}, historical {[}0.28, 0.72{]}; the groups overlap only at the
boundary, where the low end of the neoclassical group (Frahm,
\(R^2 = 0.56\)) meets the high end of the historical group (Mendelssohn,
\(R^2 = 0.72\)). The slope exponents \(\alpha\) are comparable between
groups (neoclassical mean 1.39, historical mean 1.42), indicating that
what distinguishes neoclassical from historical is not \emph{how steep}
the rank-frequency curve is but \emph{how cleanly it is a power law}.

An intuitive reading: a high \(R^2\) on a rank-frequency fit means that
the distribution has a sharp hierarchy : a small set of transitions used
very often, a long tail used rarely, and a smooth decay between. In the
neoclassical artists we examined, this hierarchy is substantially more
regular than in any of the historical composers we analyzed. This is
consistent with a defining feature of minimalist compositional practice:
a compact harmonic vocabulary deployed with systematic repetition. We
emphasize that this observation pertains to the five artists examined
and that its generalization to minimalism more broadly (symphonic
minimalism, American minimalist tradition, or earlier minimalist
precedents such as Satie) is an empirical question beyond the scope of
this study.

Additional quantities : KL divergence between individual neoclassical
artists and historical era pools, sensitivity of these assignments to
pool subsampling, comparison with cosine similarity on hand-engineered
feature vectors : are reported in Appendix C.

\begin{center}\rule{0.5\linewidth}{0.5pt}\end{center}

\subsection{6. Discussion}\label{discussion}

We discuss three topics: methodological limitations that bound the
interpretation of our results (Section 6.1), what the measures do and do
not capture about musical style (Section 6.2), and implications for
future work (Section 6.3). We deliberately avoid strong claims and
signal explicitly where our findings are robust, where they are
suggestive, and where they are contingent on methodological choices that
could have been made otherwise.

\subsubsection{6.1 Methodological
limitations}\label{methodological-limitations}

\textbf{Transcription error propagation.} The information-theoretic
measures reported here operate on symbolic distributions derived from
automatic transcription. Even at a certified F1 of 0.979 on MAESTRO, the
residual error rate is non-zero; it varies systematically by era, with
late-Romantic and Impressionist repertoire (Liszt, Scriabin,
Rachmaninoff) transcribed at 0.958--0.970 F1 versus 0.989--0.995 for
Baroque and Classical. Transcription errors are unlikely to be
distributed uniformly across degree categories; dense chromatic passages
are both harder to transcribe accurately and likely to generate spurious
events on altered degrees that inflate the tail of the scale-degree
distribution. We have not attempted to correct for this bias. The
Shannon and Zipf measures on late-Romantic composers in particular
should therefore be read as upper bounds on harmonic vocabulary breadth:
some of the apparent complexity may reflect transcription noise rather
than compositional intent.

\textbf{Tonic estimation.} Tonic detection (Section 3.5) is performed by
Essentia's \texttt{KeyExtractor} with the \texttt{bgate} profile of
Faraldo et al.~{[}2016{]}. While Faraldo et al.~report that
\texttt{bgate} outperforms the classical Krumhansl--Kessler profile on
several tonal corpora, the profile was originally derived from
electronic dance music annotations, and its behaviour on highly
chromatic or modulating classical repertoire has not been independently
validated in the form it is used here. Because all subsequent measures
operate on degrees \emph{relative to the estimated tonic}, errors in
tonic estimation propagate multiplicatively: a per-piece tonic error
shifts the entire scale-degree stream for that piece and distorts both
its marginal and its transition distribution. We have not quantified
per-composer tonic-error rates. A systematic sensitivity analysis
against alternative key-finding algorithms (e.g., Temperley's
probabilistic model, the Albrecht--Shanahan profile) is a natural
extension we leave to future work; the absolute values of Shannon, KL,
and Zipf reported here should therefore be read with this caveat, even
though rank-order comparisons between composers are likely more stable
than absolute values under such errors.

\textbf{Composer--performer confounding.} Our measures aggregate across
the multiple performers who appear in MAESTRO for each composer. The
information-theoretic measures reported in this paper are computed on
\emph{harmonic scale-degree distributions} : quantities that reflect
note choice rather than expressive interpretation : and are therefore
largely insensitive to this confound. We do not claim complete
independence from performer choice, but the confound is weaker for this
layer than for expressive metrics (velocity, rubato, articulation) where
previous cross-performer validation has shown substantial
performer-attributable variance. One residual performer-dependent effect
is worth noting: the transcription model preserves sustain-pedal events
{[}Kong et al.~2021{]}, and pedaling choices that bridge or separate
adjacent harmonies can shift the boundaries Essentia's
\texttt{ChordsDetection} places between successive chords, affecting the
transition distribution (though not the marginal). This effect is more
likely to matter for densely pedaled late-Romantic repertoire than for
the sparse pedaling typical of Baroque and Classical performance
traditions; a dedicated sensitivity analysis against pedal-stripped
transcriptions is left to future work.

\textbf{Alphabet choice.} The \textbar 𝒟\textbar{} = 15 scale-degree
alphabet is a design choice with specific consequences. It collapses all
chord qualities within a given root-to-tonic relationship (major and
minor I/IV/V distinguished; seventh chords, extended qualities, and
inversions not) and treats each degree change as a single event
regardless of duration or metric position. The choice was driven by the
requirement of statistical stability across the full corpus: with the
smallest sub-corpus at N = 11 pieces (Philip Glass) and Laplace
smoothing at α = 0.5, a larger alphabet incorporating chord qualities
would have produced a 50+ symbol space in which most cells receive zero
raw counts and the smoothed distribution is dominated by the prior
rather than by the data. A function-aware alphabet distinguishing V from
V7, ii from ii°, or IV from IV maj7 would be preferable in principle :
it would carry more harmonic information per symbol : but requires
larger per-composer sample sizes than those available for the
neoclassical sub-corpus. Different alphabet designs (e.g., the
chord-type alphabet of Serrà et al.~{[}2019{]} on larger corpora, or a
smaller root-class alphabet) would produce different absolute values
while likely preserving the qualitative rankings for the measures
reported here. The low discriminative power of Shannon entropy on this
alphabet (Section 5.1) is partly a consequence of this choice: a richer
alphabet would spread composers over a wider entropy range. The KL and
Zipf measures, which operate on pairwise comparisons and rank-frequency
shapes respectively, are less sensitive to the alphabet size.

\textbf{Corpus composition.} The MAESTRO v3.0.0 corpus is drawn from
competition performances and is not a balanced random sample of the
piano repertoire: composer representation ranges from three to 198
pieces, and era representation is dominated by Romantic repertoire (391
pieces versus 69 for Late-Romantic, for example). Our measures are
normalized per composer, so the sample-size imbalance does not directly
bias pairwise comparisons; it does, however, mean that some composer
estimates have wider bootstrap confidence intervals than others
(Appendix A), and readers interpreting the KL matrix should give more
weight to pairs involving large-sample composers.

\textbf{Sample size for per-composer measures.} Thirteen composers in
MAESTRO are represented by fewer than ten pieces (Berg, Grieg,
Mendelssohn/Rachmaninoff, Balakirev, Janáček, Franck, Handel,
Bach/Busoni, Medtner, Clementi, Tchaikovsky, Tchaikovsky/Pletnev,
Albéniz). These are excluded from the main per-composer analyses (Tables
2, 3, 4); their individual profiles are reported in Appendix A with
explicit low-sample flags. A subset of them contributes to the era-level
aggregates of Appendix C, where pooling is used as an exploratory device
rather than a primary finding: we also report there (Section C.2) a
subsampling analysis showing that the specific within-pre-Romantic
neoclassical-to-era assignments are sensitive to pool composition, which
is one reason we elevate the Zipfian-regularity finding (Section 5.4)
rather than the era-pool assignments as the robust corpus-level result.
Readers interested in the low-sample composers should note the wide
bootstrap confidence intervals.

\subsubsection{6.2 What the measures capture, and what they do
not}\label{what-the-measures-capture-and-what-they-do-not}

\textbf{Shannon entropy on the marginal.} The narrow range of 3.33 to
3.86 bits across high-sample composers (Section 5.1) reflects a real
feature of tonal piano repertoire rather than a measurement artifact: at
the level of which degrees appear, tonal composers use broadly similar
vocabularies. Differences emerge in \emph{how frequently} each degree is
used, and especially in \emph{what transitions between degrees} are
preferred : which is why the KL and Zipf measures on the full transition
matrix are substantially more discriminative than the marginal Shannon.
Shannon is reported in this paper primarily as a descriptor of the
overall marginal shape, not as a primary tool for stylistic separation.

\textbf{KL divergence on the marginal.} The pairs recovered in Table 3
(Haydn--Beethoven, Liszt--Rachmaninoff, Schubert--Schumann) are
stylistic lineages well established in musicology and music-theoretic
analysis. The fact that they emerge from a distributional comparison of
automatically transcribed audio, with no labels and no supervised
training, provides indirect evidence that the transcription pipeline
preserves enough harmonic signal to support this kind of analysis. The
KL matrix should \emph{not} be read as a complete model of style: it
captures only the scale-degree marginal, and two composers with similar
marginals may still differ substantially in rhythm, phrasing, form, and
the sequential structure of degree use.

\textbf{Zipfian fits.} The Zipfian gap reported in Section 5.4 : mean
\(R^2 = 0.78\) for neoclassical versus 0.46 for historical : captures
the degree to which a composer's transition distribution follows a clean
rank-frequency law. A high \(R^2\) indicates a sharper hierarchical
structure: a small number of transitions used very often, a long tail
used rarely, and a smooth decay in between. A lower \(R^2\) indicates
that the decay is noisier, i.e., the composer's transition vocabulary is
more evenly distributed across ranks. The finding that neoclassical
artists fit Zipfian laws substantially better than historical composers
do is, in this sense, consistent with the minimalist aesthetic: a
compact vocabulary repeated with frequency-rank regularity. We emphasize
that this is a \emph{corpus-level} observation on the five artists we
examined; whether it generalizes to minimalism more broadly : Glass's
symphonic work, Reich, Riley, Adams, or to unrelated minimalist
traditions such as Feldman or Young : is a question we do not address.

\textbf{What we do not measure.} We have not attempted to measure voice
leading, motivic development, tonal tension in the sense of
Lerdahl--Jackendoff, or hierarchical (Schenkerian) structure. The
information-theoretic profile of a composer captures a horizontal
harmonic vocabulary; it is silent on contrapuntal logic, narrative
shape, and the kind of large-scale organization that constitutes a
substantial part of what musicologists mean by ``style''. Readers should
interpret the distance and similarity claims of Section 5 accordingly:
two composers close in KL may still produce radically different music at
every level above the harmonic marginal.

\subsubsection{6.3 Implications and future
work}\label{implications-and-future-work}

The pipeline described in this paper makes three things possible that
were previously impractical at corpus scale: (i) reproducing
symbolic-level musicological measures from audio alone, with no score or
MIDI input; (ii) extending such measures beyond curated symbolic corpora
to include any piano recording that enters the pipeline, including
living composers and genres poorly represented in symbolic databases;
(iii) grounding corpus-scale stylistic comparison in quantifiable
information-theoretic objects with explicit confidence intervals.

The neoclassical case study illustrates each point. No large symbolic
dataset of Richter, Frahm, Glass, Arnalds, or Jóhannsson exists at the
scale needed for reliable per-composer distributional analysis; the
pipeline produces one directly from commercial audio. The resulting
profiles sit alongside the MAESTRO profiles in a common mathematical
space, allowing comparative questions : is the neoclassical transition
vocabulary more Zipfian than Bach's? : to be asked and answered rather
than argued. Whether the specific answers reported here survive
replication on larger and more diverse contemporary corpora is an
empirical question that this paper is not in a position to settle.

Several natural extensions suggest themselves. The harmonic layer could
be enriched to include chord qualities, inversions, and voice-leading
proxies; this would lift some of the alphabet-choice limitations
discussed above. Additional contemporary sub-corpora (film scores, jazz
transcriptions, non-Western piano traditions) could be placed on the
same information-theoretic map, testing whether the Zipfian regularity
reported here is specific to these five artists or characteristic of a
broader stylistic space. Cross-modal comparisons : for example, between
the information-theoretic profile of a composer's piano output and their
orchestral output : would require extending the transcription layer
beyond piano, but could eventually place the full repertoire in a single
quantitative framework. We leave these to future work.

\begin{center}\rule{0.5\linewidth}{0.5pt}\end{center}

\subsection{7. Conclusion}\label{conclusion}

We have described an end-to-end pipeline (Cygnus Analysis) that takes
raw piano audio as input and produces per-composer information-theoretic
profiles at corpus scale, with certified transcription accuracy on a
standard MIR benchmark. Applied to 1,238 pieces across 15 MAESTRO
composers with at least ten attributed pieces in MAESTRO v3.0.0 (plus
thirteen low-sample composers reported separately in Appendix A), the
pipeline yields three main results.

First, Shannon entropy on the 15-symbol scale-degree alphabet spans a
narrow range of 3.33 to 3.86 bits across composers, with the upper
cluster occupying 97--99\% of the uniform bound. Marginal entropy alone
is therefore a weak discriminator of tonal style at this alphabet size;
the discriminative signal lives at the level of degree combinations and
frequencies, not of degree presence.

Second, asymmetric Kullback--Leibler divergence on the same
distributions recovers documented stylistic lineages without
supervision. The smallest pairwise divergences in the corpus
(Haydn--Beethoven, Liszt--Rachmaninoff, Schubert--Schumann) correspond
to relationships described by music-historical scholarship. Mendelssohn
emerges as a stable outlier at the marginal level, with the largest
divergences in the corpus running between his profile and those of his
Classical and early-Romantic peers. Results are robust to the choice of
Laplace smoothing parameter across \(\alpha \in \{0.1, 0.5, 1.0\}\).

Third, Zipfian rank-frequency fits on the 15 × 15 transition
distribution separate the five contemporary neoclassical artists we
examined (Richter, Frahm, Glass, Arnalds, Jóhannsson) from the fifteen
MAESTRO composers with N ≥ 10, with mean \(R^2 = 0.78\) versus 0.46.
This gap is larger than the spread within either group and is stable
under bootstrap resampling and alternative Laplace smoothing parameters.
It reflects a distinctive feature of the neoclassical transition
vocabulary: a compact set of transitions deployed with sharper
frequency-rank regularity than in any of the historical composers we
analyze. The slope exponent \(\alpha\) is comparable between groups
(\textasciitilde1.4); what separates them is the \emph{quality of the
power-law fit}, not its steepness.

The contribution of this paper is neither a new information-theoretic
technique nor a new transcription architecture. It is an integrated
demonstration that certified audio transcription combined with classical
information-theoretic analysis can produce a coherent, corpus-scale,
reproducible map of the piano repertoire on which historical and
contemporary composers can be jointly located, and that such a map
exposes quantitative regularities : some expected, some unexpected :
that are worth the attention of the musicology and MIR communities. Our
data, confidence intervals, and intermediate artifacts are released
alongside this paper to support independent replication and extension.

\begin{center}\rule{0.5\linewidth}{0.5pt}\end{center}

\subsection{Ethics and data statement}\label{ethics-and-data-statement}

The MAESTRO v3.0.0 corpus is released under a CC BY-NC-SA 4.0 license
for non-commercial research use, under which the primary analyses of
this paper are conducted. The 111 contemporary neoclassical recordings
analyzed in Section 5.5 are publicly available commercial releases used
exclusively for non-commercial academic research under applicable
fair-dealing and fair-use provisions. No audio excerpts, derived audio,
or full MIDI transcriptions of the neoclassical recordings are
redistributed with this paper or in the supplementary materials. Only
aggregate distributional statistics : per-composer scale-degree count
vectors, KL and Jensen--Shannon divergence matrices, Shannon entropies,
Zipfian fit parameters, and bootstrap confidence intervals : are
released. This ensures non-reproducibility of the original copyrighted
audio from the supplementary materials alone. For the MAESTRO subset, we
release transcribed MIDI only where already permitted under the CC
BY-NC-SA 4.0 license of the source dataset.

\begin{center}\rule{0.5\linewidth}{0.5pt}\end{center}

\subsection{Appendix A: Low-sample composers (N \textless{} 10
pieces)}\label{appendix-a-low-sample-composers-n-10-pieces}

Thirteen composers in MAESTRO v3.0.0 are represented by fewer than ten
pieces and are excluded from the main per-composer analyses of Section 5
(Shannon ranking, KL matrix, Zipfian fits). Some are nevertheless pooled
into the era aggregates of Appendix C to preserve era coverage. We
report their information-theoretic profiles here for completeness, with
the understanding that bootstrap confidence intervals are substantially
wider than for high-sample composers.

\textbf{Table A.1}: Information-theoretic profile for low-sample
composers (N \textless{} 10 pieces), ordered by N.

{\def\LTcaptype{none} 
\begin{longtable}[]{@{}
  >{\raggedright\arraybackslash}p{(\linewidth - 10\tabcolsep) * \real{0.1304}}
  >{\raggedleft\arraybackslash}p{(\linewidth - 10\tabcolsep) * \real{0.1739}}
  >{\raggedleft\arraybackslash}p{(\linewidth - 10\tabcolsep) * \real{0.1739}}
  >{\raggedleft\arraybackslash}p{(\linewidth - 10\tabcolsep) * \real{0.1739}}
  >{\raggedleft\arraybackslash}p{(\linewidth - 10\tabcolsep) * \real{0.1739}}
  >{\raggedleft\arraybackslash}p{(\linewidth - 10\tabcolsep) * \real{0.1739}}@{}}
\toprule\noalign{}
\begin{minipage}[b]{\linewidth}\raggedright
Composer
\end{minipage} & \begin{minipage}[b]{\linewidth}\raggedleft
N
\end{minipage} & \begin{minipage}[b]{\linewidth}\raggedleft
Shannon \(H\) (bits)
\end{minipage} & \begin{minipage}[b]{\linewidth}\raggedleft
95\% CI
\end{minipage} & \begin{minipage}[b]{\linewidth}\raggedleft
Zipf \(\alpha\) (trans.)
\end{minipage} & \begin{minipage}[b]{\linewidth}\raggedleft
Zipf \(R^2\) (trans.)
\end{minipage} \\
\midrule\noalign{}
\endhead
\bottomrule\noalign{}
\endlastfoot
Alban Berg & 3 & 3.519 & {[}3.511, 3.522{]} & 1.416 & 0.672 \\
Edvard Grieg & 3 & 3.366 & {[}2.993, 3.502{]} & 0.951 & 0.866 \\
Felix Mendelssohn / Sergei Rachmaninoff & 3 & 3.423 & {[}3.338, 3.432{]}
& 1.319 & 0.797 \\
Mily Balakirev & 4 & 3.684 & {[}3.461, 3.812{]} & 1.230 & 0.725 \\
Leoš Janáček & 4 & 3.679 & {[}3.300, 3.770{]} & 1.101 & 0.706 \\
César Franck & 5 & 3.725 & {[}3.360, 3.768{]} & 1.166 & 0.569 \\
George Frideric Handel & 5 & 3.688 & {[}3.084, 3.752{]} & 1.402 &
0.718 \\
Johann Sebastian Bach / Ferruccio Busoni & 5 & 3.466 & {[}3.270,
3.742{]} & 1.588 & 0.793 \\
Nikolai Medtner & 5 & 3.664 & {[}3.432, 3.824{]} & 1.212 & 0.654 \\
Muzio Clementi & 6 & 3.668 & {[}3.254, 3.753{]} & 1.428 & 0.739 \\
Pyotr Ilyich Tchaikovsky & 6 & 3.599 & {[}3.170, 3.773{]} & 1.298 &
0.757 \\
Pyotr Ilyich Tchaikovsky / Mikhail Pletnev & 6 & 3.536 & {[}3.212,
3.791{]} & 1.487 & 0.808 \\
Isaac Albéniz & 7 & 3.750 & {[}3.447, 3.829{]} & 1.173 & 0.632 \\
\end{longtable}
}

\emph{Note}: Zipf fits were computed for all thirteen low-sample
composers on the 225-point transition distribution using the same
protocol as Section 4.5. The wide confidence intervals for Grieg,
Handel, and several others reflect the small number of pieces available.
These profiles should not be used for direct comparison with the
high-sample composers of Tables 2--4 without explicit acknowledgment of
the sample-size discrepancy.

\begin{center}\rule{0.5\linewidth}{0.5pt}\end{center}

\subsection{Appendix B: Full KL divergence
matrix}\label{appendix-b-full-kl-divergence-matrix}

The complete asymmetric KL divergence matrix for the 20 high-sample
entities (15 MAESTRO composers with N ≥ 10 plus the 5 neoclassical
artists) is visualized in Figure B.1. The full 33 × 33 matrix including
the thirteen low-sample composers is released as
\texttt{kl\_scale\_degrees\_33x33.json}, with bootstrap 95\% confidence
intervals in the same file (\texttt{bootstrap\_ci} field). The analogous
matrix on scale-degree transitions is in
\texttt{kl\_transitions\_33x33.json}; the symmetric Jensen--Shannon
cross-check is in \texttt{js\_scale\_degrees\_33x33.json}.

\textbf{Figure B.1}: Asymmetric Kullback--Leibler divergence
\(D_{\mathrm{KL}}(P_a \parallel P_b)\) between scale-degree marginals
for the 20 high-sample entities. Rows = source composer, columns =
target; darker cells indicate \emph{larger} divergence (magma\_r
colormap). Composers are grouped by era along both axes (Baroque,
Classical, Classical--Romantic, Romantic, Late-Romantic, Impressionist,
Neoclassical); the neoclassical block is highlighted with a dashed teal
rectangle. Values are capped at 1.5 bits for visualization so that finer
contrasts among the non-Mendelssohn composers remain legible; the
maximum observed value across the matrix is 2.76 bits (Mozart ‖
Mendelssohn). Laplace smoothing \(\alpha = 0.5\), log base 2.

\begin{figure}
\centering
\pandocbounded{\includegraphics[keepaspectratio,alt={Figure B.1: KL heatmap}]{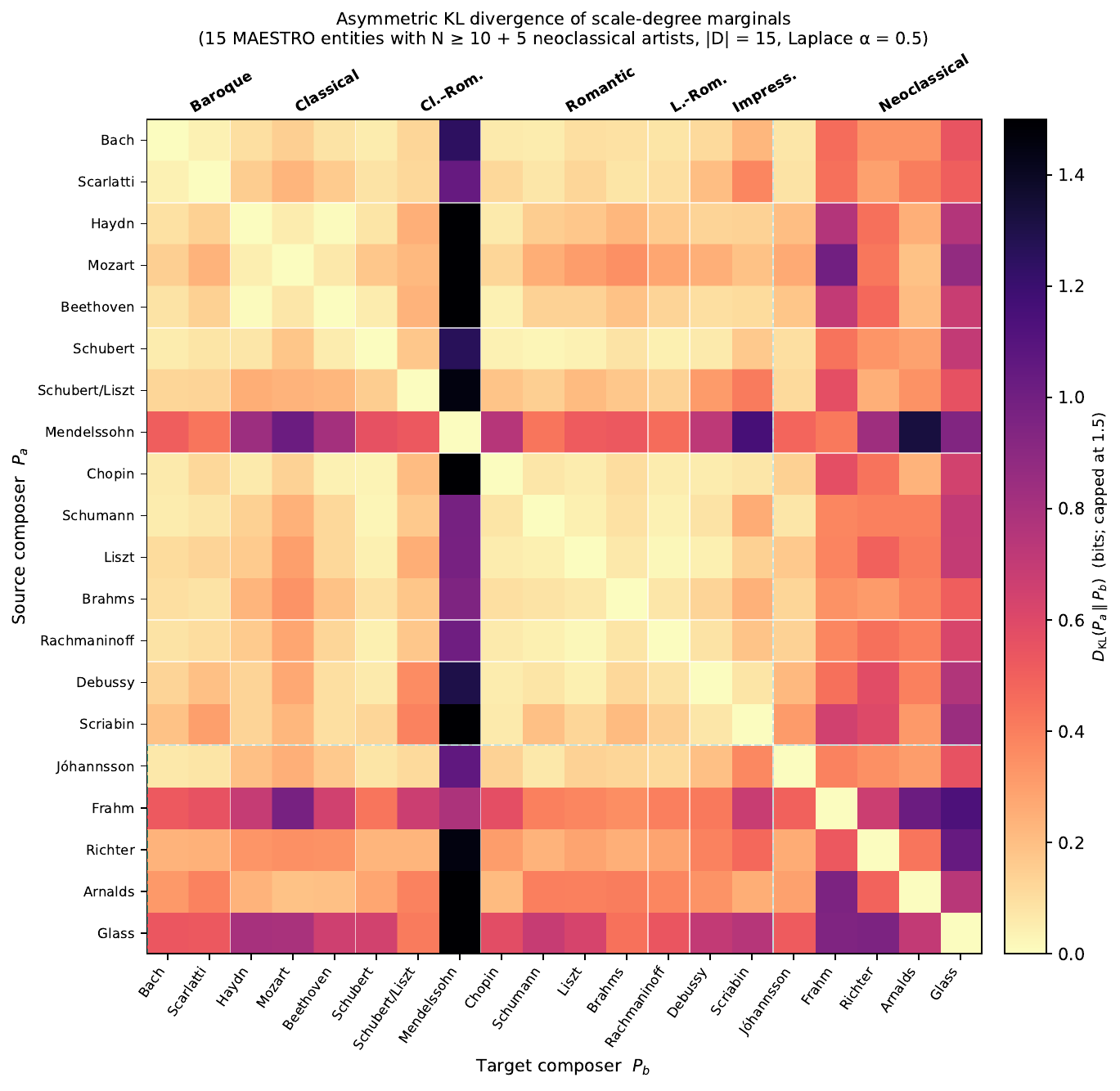}}
\caption{Figure B.1: KL heatmap}
\end{figure}

\begin{center}\rule{0.5\linewidth}{0.5pt}\end{center}

\subsection{Appendix C: Neoclassical
extensions}\label{appendix-c-neoclassical-extensions}

This appendix reports additional analyses of the five neoclassical
artists that were examined during the study but not retained as primary
findings in Section 5. These results are presented for transparency and
to support follow-up investigation.

\subsubsection{C.1 KL divergence against era
pools}\label{c.1-kl-divergence-against-era-pools}

The KL divergence of each neoclassical artist against each historical
era pool is reported in Table C.1. Era pools are defined using the N ≥
10 threshold: Baroque (Scarlatti, Handel, Bach, Bach/Busoni
transcription), Classical (Haydn, Mozart, Beethoven), Classical-Romantic
(Schubert, Schubert/Liszt transcription, Mendelssohn,
Mendelssohn/Rachmaninoff transcription), Romantic (Chopin, Schumann,
Liszt, Brahms), Late-Romantic (Rachmaninoff, Grieg, Albéniz), and
Impressionist (Debussy, Scriabin, Janáček). Pool distributions are
constructed by concatenating raw counts across all constituent
composers, then applying Laplace α = 0.5 smoothing.

\textbf{Table C.1}:
\(D_{\mathrm{KL}}(P_{\text{artist}} \parallel P_{\text{era}})\) in bits,
scale-degree marginal.

{\def\LTcaptype{none} 
\begin{longtable}[]{@{}
  >{\raggedright\arraybackslash}p{(\linewidth - 12\tabcolsep) * \real{0.1111}}
  >{\raggedleft\arraybackslash}p{(\linewidth - 12\tabcolsep) * \real{0.1481}}
  >{\raggedleft\arraybackslash}p{(\linewidth - 12\tabcolsep) * \real{0.1481}}
  >{\raggedleft\arraybackslash}p{(\linewidth - 12\tabcolsep) * \real{0.1481}}
  >{\raggedleft\arraybackslash}p{(\linewidth - 12\tabcolsep) * \real{0.1481}}
  >{\raggedleft\arraybackslash}p{(\linewidth - 12\tabcolsep) * \real{0.1481}}
  >{\raggedleft\arraybackslash}p{(\linewidth - 12\tabcolsep) * \real{0.1481}}@{}}
\toprule\noalign{}
\begin{minipage}[b]{\linewidth}\raggedright
Artist
\end{minipage} & \begin{minipage}[b]{\linewidth}\raggedleft
Baroque
\end{minipage} & \begin{minipage}[b]{\linewidth}\raggedleft
Classical
\end{minipage} & \begin{minipage}[b]{\linewidth}\raggedleft
Cl.-Rom.
\end{minipage} & \begin{minipage}[b]{\linewidth}\raggedleft
Romantic
\end{minipage} & \begin{minipage}[b]{\linewidth}\raggedleft
Late-Rom.
\end{minipage} & \begin{minipage}[b]{\linewidth}\raggedleft
Imp.
\end{minipage} \\
\midrule\noalign{}
\endhead
\bottomrule\noalign{}
\endlastfoot
Jóhannsson & \textbf{0.061} & 0.175 & 0.063 & 0.096 & 0.098 & 0.214 \\
Richter & 0.231 & 0.334 & \textbf{0.226} & 0.255 & 0.267 & 0.385 \\
Frahm & 0.519 & 0.691 & \textbf{0.384} & 0.422 & 0.381 & 0.476 \\
Arnalds & 0.333 & \textbf{0.189} & 0.371 & 0.306 & 0.386 & 0.286 \\
Glass & \textbf{0.539} & 0.682 & 0.609 & 0.574 & 0.564 & 0.693 \\
\end{longtable}
}

On the marginal, three of five artists are KL-closest to pre-Romantic
pools (Baroque, Classical, or Classical-Romantic). The same pattern
holds on the scale-degree transition distribution (Supplementary data).

\subsubsection{C.2 Sensitivity to pool
size}\label{c.2-sensitivity-to-pool-size}

Because the era pools are unequal in size (Romantic N = 391,
Late-Romantic N = 69), we recomputed Table C.1 with all
non-Late-Romantic pools subsampled to 69 pieces (100 random subsamples
per pool, seeds 42--141, without replacement, with Late-Romantic kept at
full size). Concordance rates between the subsampled min-era assignments
and the full-sample assignments of Table C.1 are reported in Table C.2.

\textbf{Table C.2}: Concordance of subsampled min-era with full-sample
min-era (100 subsamples each).

{\def\LTcaptype{none} 
\begin{longtable}[]{@{}
  >{\raggedright\arraybackslash}p{(\linewidth - 6\tabcolsep) * \real{0.2143}}
  >{\raggedright\arraybackslash}p{(\linewidth - 6\tabcolsep) * \real{0.2143}}
  >{\raggedleft\arraybackslash}p{(\linewidth - 6\tabcolsep) * \real{0.2857}}
  >{\raggedleft\arraybackslash}p{(\linewidth - 6\tabcolsep) * \real{0.2857}}@{}}
\toprule\noalign{}
\begin{minipage}[b]{\linewidth}\raggedright
Artist
\end{minipage} & \begin{minipage}[b]{\linewidth}\raggedright
Full-sample min-era
\end{minipage} & \begin{minipage}[b]{\linewidth}\raggedleft
Concordance (marginal)
\end{minipage} & \begin{minipage}[b]{\linewidth}\raggedleft
Concordance (transitions)
\end{minipage} \\
\midrule\noalign{}
\endhead
\bottomrule\noalign{}
\endlastfoot
Jóhannsson & Baroque & 87\% & \textbf{100\%} \\
Richter & Baroque & 65\% & \textbf{96\%} \\
Frahm & Classical-Romantic & 76\% & 40\% \\
Arnalds & Classical & \textbf{99\%} & 89\% \\
Glass & Baroque & 53\% & 78\% \\
\end{longtable}
}

Two assignments remain robust (≥ 95\% concordance) on at least one
metric: Jóhannsson → Baroque on transitions (100\%), and Arnalds →
Classical on marginal (99\%). The remaining assignments are sensitive to
pool sampling, typically flipping to adjacent pools within the
pre-Romantic / Classical-Romantic range. Across all 500 subsamples ×
five artists × two metrics, no subsample assigns Romantic or
Impressionist as the closest era for Jóhannsson, Arnalds, Richter, or
Glass; Frahm's assignment occasionally flips to Romantic or
Late-Romantic. \textbf{The coarser observation that neoclassical artists
tend to fall outside the Romantic / Impressionist range is preserved
under subsampling; the specific within-pre-Romantic assignment is not.}

This sensitivity is the reason we do not retain the specific
``neoclassical → Baroque'' assignments as a primary finding in Section
5, and instead elevate the more robust Zipfian-regularity finding.

\subsubsection{C.3 Comparison with cosine
similarity}\label{c.3-comparison-with-cosine-similarity}

A parallel analysis using cosine similarity on the hand-engineered
feature vectors of the Cygnus profiler (entropy, Gini, chromaticism,
velocity, density, rubato, pedal events, structural climax position)
produces different nearest-neighbor assignments for several artists.
Spearman rank correlation between the KL ranking and the cosine ranking,
computed per neoclassical artist over the 23 historical composers
covered by both comparators, is reported in Table C.3.

\textbf{Table C.3}: Spearman \(\rho\) between KL and cosine rankings of
historical neighbors.

{\def\LTcaptype{none} 
\begin{longtable}[]{@{}lrr@{}}
\toprule\noalign{}
Artist & Spearman \(\rho\) & p-value \\
\midrule\noalign{}
\endhead
\bottomrule\noalign{}
\endlastfoot
Max Richter & 0.601 & 0.002 \\
Jóhann Jóhannsson & 0.490 & 0.015 \\
Philip Glass & 0.308 & 0.143 \\
Ólafur Arnalds & 0.170 & 0.426 \\
Nils Frahm & 0.152 & 0.478 \\
\end{longtable}
}

The two rankings agree moderately for Richter and Jóhannsson and
disagree substantially for the other three artists. The divergence
reflects the different objects the two measures operate on: cosine on a
hand-engineered feature vector mixes harmonic, rhythmic, and expressive
dimensions; KL on the scale-degree marginal isolates the harmonic layer.
We do not attempt to arbitrate between the two measures here. Detailed
disagreement patterns are in supplementary data
(\texttt{sanity\_check\_kl\_vs\_cosine.md}).

\begin{center}\rule{0.5\linewidth}{0.5pt}\end{center}

\subsection{References}\label{references}

\begin{enumerate}
\def\labelenumi{\arabic{enumi}.}
\item
  \textbf{Agresti, A., \& Coull, B. A.} (1998). Approximate is better
  than ``exact'' for interval estimation of binomial proportions.
  \emph{The American Statistician}, 52(2), 119--126.
\item
  \textbf{Bogdanov, D., et al.} (2013). Essentia: An Audio Analysis
  Library for Music Information Retrieval. \emph{ISMIR 2013}.
\item
  \textbf{Bradshaw, L., et al.} (2025). Aria-MIDI: A Dataset of Piano
  MIDI Files for Symbolic Music Modeling. \emph{ICLR 2025}.
\item
  \textbf{Cover, T. M., \& Thomas, J. A.} (2006). \emph{Elements of
  Information Theory} (2nd ed.). Wiley.
\item
  \textbf{Cuthbert, M. S., \& Ariza, C.} (2010). music21: A toolkit for
  computer-aided musicology and symbolic music data. \emph{ISMIR 2010}.
\item
  \textbf{Febres, G., \& Jaffé, K.} (2017). Music viewed by its entropy
  content: A novel window for comparative analysis. \emph{PLOS ONE},
  12(10), e0185757.
\item
  \textbf{Hawthorne, C., Elsen, E., Song, J., Roberts, A., Simon, I.,
  Raffel, C., Engel, J., Oore, S., \& Eck, D.} (2018). Onsets and
  Frames: Dual-Objective Piano Transcription. \emph{ISMIR 2018}.
\item
  \textbf{Hawthorne, C., Stasyuk, A., Roberts, A., Simon, I., Huang,
  C.-Z. A., Dieleman, S., Elsen, E., Engel, J., \& Eck, D.} (2019).
  Enabling Factorized Piano Music Modeling and Generation with the
  MAESTRO Dataset. \emph{ICLR 2019}.
\item
  \textbf{Knopoff, L., \& Hutchinson, W.} (1981). Information Theory for
  Musical Continua. \emph{Journal of Music Theory}, 25(1), 17--44.
\item
  \textbf{Knopoff, L., \& Hutchinson, W.} (1983). Entropy as a Measure
  of Style: The Influence of Sample Length. \emph{Journal of Music
  Theory}, 27(1), 75--97.
\item
  \textbf{Kong, Q., Li, B., Song, X., Wan, Y., \& Wang, Y.} (2021).
  High-resolution Piano Transcription with Pedals by Regressing Onset
  and Offset Times. \emph{IEEE/ACM Transactions on Audio, Speech, and
  Language Processing}, 29, 3707--3717.
\item
  \textbf{Kong, Q., et al.} (2020). GiantMIDI-Piano: A large-scale MIDI
  dataset for classical piano music. \emph{arXiv:2010.07061}.
\item
  \textbf{Liu, L., Wei, J., Zhang, H., Xin, J., \& Huang, J.} (2013). A
  statistical physics view of pitch fluctuations in the classical music
  from Bach to Chopin: Evidence for scaling. \emph{PLOS ONE}, 8(3),
  e58710.
\item
  \textbf{Manaris, B., Romero, J., Machado, P., Krehbiel, D., Hirzel,
  T., Pharr, W., \& Davis, R. B.} (2005). Zipf's Law, Music
  Classification, and Aesthetics. \emph{Computer Music Journal}, 29(1),
  55--69.
\item
  \textbf{McKay, C.} (2010). \emph{Automatic Music Classification with
  jSymbolic}. PhD thesis, McGill University.
\item
  \textbf{Pearce, M. T., \& Wiggins, G. A.} (2006). Expectation in
  melody: The influence of context and learning. \emph{Music
  Perception}, 23(5), 377--405.
\item
  \textbf{Raffel, C., McFee, B., Humphrey, E. J., Salamon, J., Nieto,
  O., Liang, D., \& Ellis, D. P. W.} (2014). mir\_eval: A transparent
  implementation of common MIR metrics. \emph{ISMIR 2014}.
\item
  \textbf{Sakellariou, J., Tria, F., Loreto, V., \& Pachet, F.} (2017).
  Maximum entropy models capture melodic styles. \emph{Scientific
  Reports}, 7, 9172.
\item
  \textbf{Serrà, J., Corral, Á., Boguñá, M., Haro, M., \& Arcos, J. L.}
  (2019). Zipf's law in music emerges by a natural choice of Zipfian
  units. \emph{Scientific Reports}, 9, 2646.
\item
  \textbf{Temperley, D.} (2014). Information Flow and Repetition in
  Music. \emph{Journal of Music Theory}, 58(2), 155--178.
\item
  \textbf{Voss, R. F., \& Clarke, J.} (1975). 1/f noise in music and
  speech. \emph{Nature}, 258, 317--318.
\item
  \textbf{Lu, W.-T., Wang, J.-C., Kong, Q., \& Hung, Y.-N.} (2023).
  Music Source Separation with Band-Split RoPE Transformer. \emph{Sound
  Demixing Challenge (SDX23)}. arXiv:2309.02612.
\item
  \textbf{Faraldo, Á., Jordà, S., \& Herrera, P.} (2016). A
  Multi-Profile Method for Key Estimation in EDM. \emph{AES Conference
  on Semantic Audio}, 2016.
\item
  \textbf{Weiss, C.} (2017). \emph{Computational Methods for
  Tonality-Based Style Analysis of Classical Music Audio Recordings}.
  PhD thesis, Technische Universität Ilmenau.
\item
  \textbf{Zipf, G. K.} (1949). \emph{Human Behavior and the Principle of
  Least Effort}. Addison-Wesley.
\end{enumerate}

\end{document}